\documentclass[pra,twocolumn,superscriptaddress]{revtex4}

\usepackage{amsfonts}
\usepackage{amsmath}
\usepackage{bbm}
\usepackage{graphicx}
\usepackage{bbold}
\usepackage{amssymb}
\usepackage{color}
\usepackage{subfigure}

\usepackage{mathtools}

\begin{document}
\title{The gauge coupled two-body problem in a ring}

\author{Joel Priestley}
\affiliation{SUPA, Institute of Photonics and Quantum Sciences, Heriot-Watt University, Edinburgh EH14 4AS, United Kingdom}
\author{Gerard Valent\'{\i}-Rojas}
\affiliation{SUPA, Institute of Photonics and Quantum Sciences, Heriot-Watt University, Edinburgh EH14 4AS, United Kingdom}
\author{Ewan M. Wright}
\affiliation{Wyant College of Optical Sciences, University of Arizona, Tucson, Arizona 85721, USA}
\author{Patrik \"Ohberg}
\affiliation{SUPA, Institute of Photonics and Quantum Sciences, Heriot-Watt University, Edinburgh EH14 4AS, United Kingdom}

\begin{abstract}
We study the properties of two quantum particles which are confined in a ring. The particles interact via a long-range gauge potential proportional to the distance between the particles. It is found that the two-body ground state corresponds to a state with non-zero angular momentum provided that the interaction between the particles is strong enough. In addition, the particles are correlated in the sense that depending on the interaction strength there is a propensity to be found close together or separated in the ring. We discuss the effect of measuring the position of one of the particles and thereby removing the particle from the ring, where we show that the remaining particle can be prepared in a non-dispersive state with non-zero angular momentum.  
\end{abstract}

\maketitle

\section{Introduction}

The quantum ring is in many respect the drosophila of quantum physics. It often provides a simple description of a range of phenomena which are related to the topology of states due to the periodicity of the ring. Prominent examples are states with persistent currents and the role of magnetic flux which penetrates the ring and thereby inducing angular momentum to the system. In the presence of interactions between particles strongly correlated situations can also be addressed successfully using for instance the Bethe ansatz.  

The basic premise in this paper is a simple one. We will discuss the properties of two gauge coupled particles in a ring, where each particle is subject to a gauge potential whose strength depends on the position of the other particle in the ring. The motivation for studying this problem is two-fold: firstly we are interested in what kind of ground states we get as a function of interaction strength. We can in particular expect a non-trivial dependence on center of mass momentum due to the gauge potential present. This question and situation is also closely related to the work by Aglietti et al. \cite{aglietti_1996} where they studied among other things the linear case with short range interactions. Secondly, a gauge potential which depends on the position of the other particle allows us to draw parallels with density dependent gauge potentials  in the many particle mean field limit \cite{edmonds2013a} which are encountered in Chern Simons theory in the context of flux attachment \cite{valenti2020}. This is further emphasised by the recent experimental realisation and emulation of a one-dimensional topological gauge theory in a Raman-coupled Bose-Einstein condensate \cite{frolian-2022} together with a comprehensive theoretical investigation of the mappings required to connect the experimental platform with the chiral BF theory \cite{chisholm-2022}.

\section{The two-body ground state \label{sec:system}}

\begin{figure}
\includegraphics[width=8.5cm]{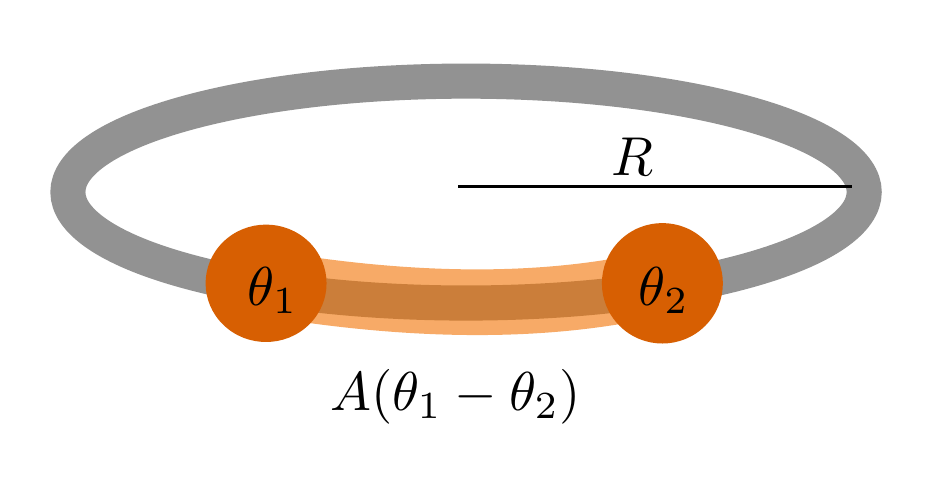}
\caption{The tightly confined ring with radius $R$ traps two particles with coordinates $\theta_1$ and $\theta_2$. The interaction between the two particles is mediated by a gauge potential $A(\theta_1-\theta_2)$ which is a function of the distance between the particles.}
\label{figexp}
\end{figure}

We envisage two particles with coordinates $x_1=R\theta_1$ and $x_2=R\theta_2$ being trapped in a ring with radius $R$ and angular coordinates $\{\theta_1,\theta_2\}=[-\pi,\pi]$, see Fig. \ref{figexp}. The two particles interact via a gauge potential in the azimuthal $\theta$-direction that is of the form $A(x_1,x_2)=\kappa\delta(x_1-x_2)/\hbar$, where $\kappa$ controls the strength of the gauge potential, and in addition we allow the $\delta(x_1-x_2)$ function to have a non-zero width, i.e. the interaction can be long range. The two-body Schr\"odinger equation for this situation is given by
\begin{eqnarray}
	i\hbar\partial_t\Psi(x_1,x_2)&=&\frac{\hbar^2}{2m}\big[(-i\partial_{x_1}+A(x_1,x_2))^2+\nonumber\\&&(-i\partial_{x_2}+A(x_1,x_2))^2\big]\Psi(x_1,x_2).
\label{sch1}
\end{eqnarray}
We assume the gauge potential $A(x_1,x_2)$ and $\delta(x_1-x_2)$ are even functions with respect to $x_1-x_2$. This in turn breaks Galilean invariance \cite{bak-1994}.  We can therefore expect chiral dynamics to appear \cite{aglietti_1996,valenti2020,valiente-a-2021,valiente-b-2021}.

It is helpful to separate variables and introduce the relative coordinate $x=x_1-x_2$ and center of mass coordinate $s=(x_1+x_2)/2$. Choosing our unit of length to be the radius $R$ of the ring and time in unit of $2mR^2/\hbar$, we obtain the transformed Schr\"odinger equation
\begin{eqnarray}
	&&i\partial_t\Psi(x,s)=\nonumber\\&& [-2\partial_{x}^2-\frac{1}{2}\partial^2_s-2i\kappa\delta(x)\partial_s+ 2\kappa^2\delta(x)^2]\Psi(x,s).
\label{sch2}
\end{eqnarray}
We write the stationary solution of Eq. (\ref{sch2}) in the form
\begin{equation}
\Psi(x,s)=e^{-i\varepsilon t} e^{ips} \varphi(x),
\end{equation}
which results in the time independent problem
\begin{equation}
[-\partial^2_x+(\frac{p}{2}+\kappa\delta(x))^2]\varphi(x)=\frac{\varepsilon}{2}\varphi(x),
\label{sch3}
\end{equation}
where $p$ is the center of mass momentum. From Eq. (\ref{sch3}) we see that the effective potential 
\begin{equation}
V_{eff}(x)=\big(\frac{p}{2}+\kappa\delta(x)\big)^2
\label{poteff}
\end{equation}
now depends on the center of mass momentum $p$ and the parameters of the gauge potential. The shape of $\delta(x)$ and the strength of $\kappa$ is fixed, and given by the physical system we happen to consider. The corresponding ground state we then find by choosing the momentum $p$ such that we obtain the lowest possible energy $\varepsilon$. From the potential in Eq. (\ref{poteff}) we can already say something about the possible ground states we can expect to see. Because of the quadratic dependence and the fact that the signs of $p$ and $\kappa$ can be different, the resulting potential can acquire the shape of a single well or a double well. This immediately tells us that there might be two kinds of solutions for the ground state, which we will show is indeed the case. 

In order to calculate explicitly what the ground state is together with its corresponding wave function we need to define the form of the gauge potential and $\delta(x)$. We will choose a localised symmetric and periodic function of the form 
\begin{equation}
\delta(x)=Q\cos^{2q}(x/2),
\label{del}
\end{equation}
where $q=1,2,3,...$ and $Q=q!/(2\sqrt{\pi}\Gamma[q+\frac{1}{2}])$ is a normalisation constant such that $\int_{-\pi}^\pi \delta(x)=1$. We note that for $q\rightarrow \infty$ we obtain the short range limit with the gauge potential proportional to the Dirac delta function. In this limit we see that the energy is given by $\varepsilon_{\ell,p}=(\ell^2+p^2)/2$ where $\ell$ is a non-zero integer and with even and odd eigenstates 
\begin{equation}
\varphi_p^\ell(x)=\left\{\begin{array}{lc}\sin(\ell\frac{|x|}{2}) &  \\\sin(\ell\frac{x}{2}) & \end{array}\right.
\end{equation}
The ground state is then given by 
$\varphi_{p=0}^{\ell=1}(x)$ with center of mass momentum $p=0$. For a long range gauge potential $\delta (x)$ with finite $q$ we will see that the situation is different. 

\begin{figure}
\includegraphics[width=8.5cm]{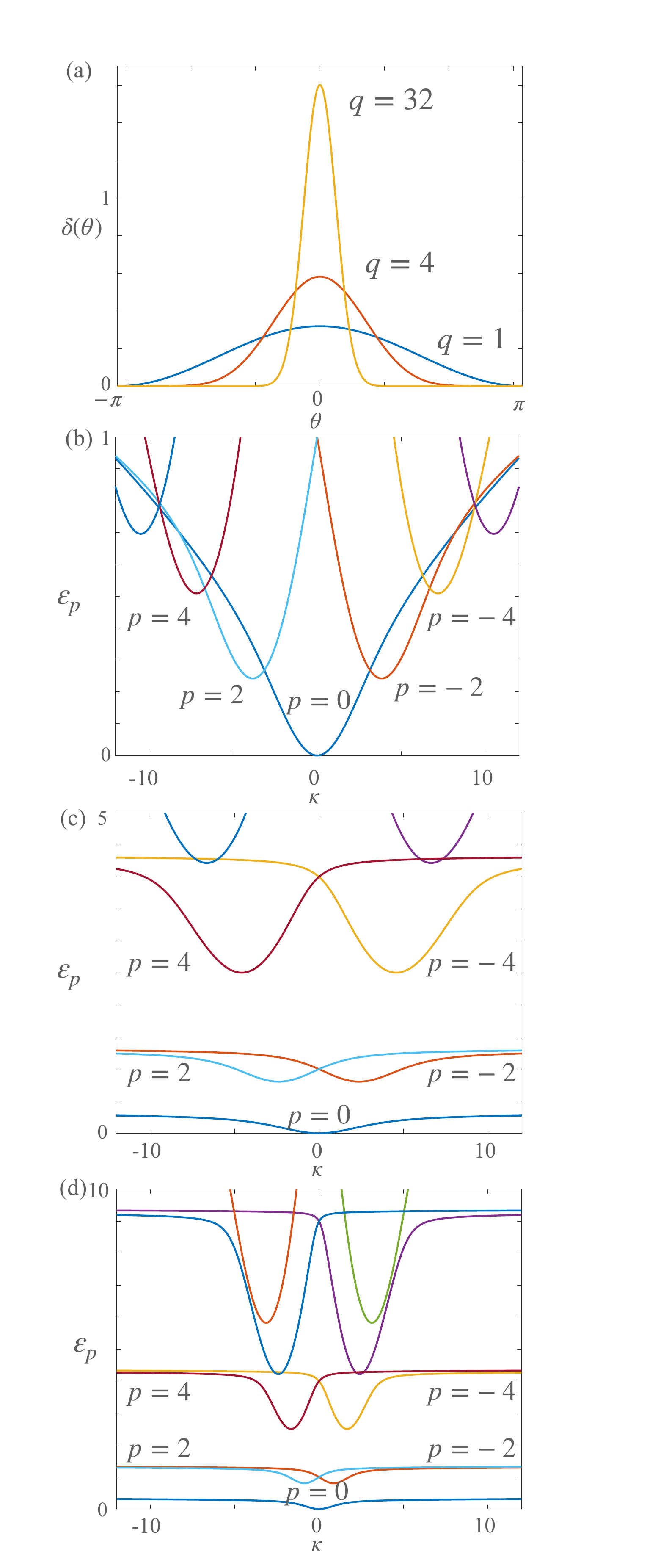}
\caption{The ground state energy as a function of $\kappa$ and for different values of the quantised centre of mass momentum $p$. Three examples of the long-range interaction are illustrated in (a). The corresponding spectra are shown in (b) for $q=1$, (c) for $q=4$, and (d) for $q=32$. With decreasing range of the interaction between the particles the system approaches the zero momentum ground state.}
\label{energy}
\end{figure}

\begin{figure}
\includegraphics[width=8.5cm]{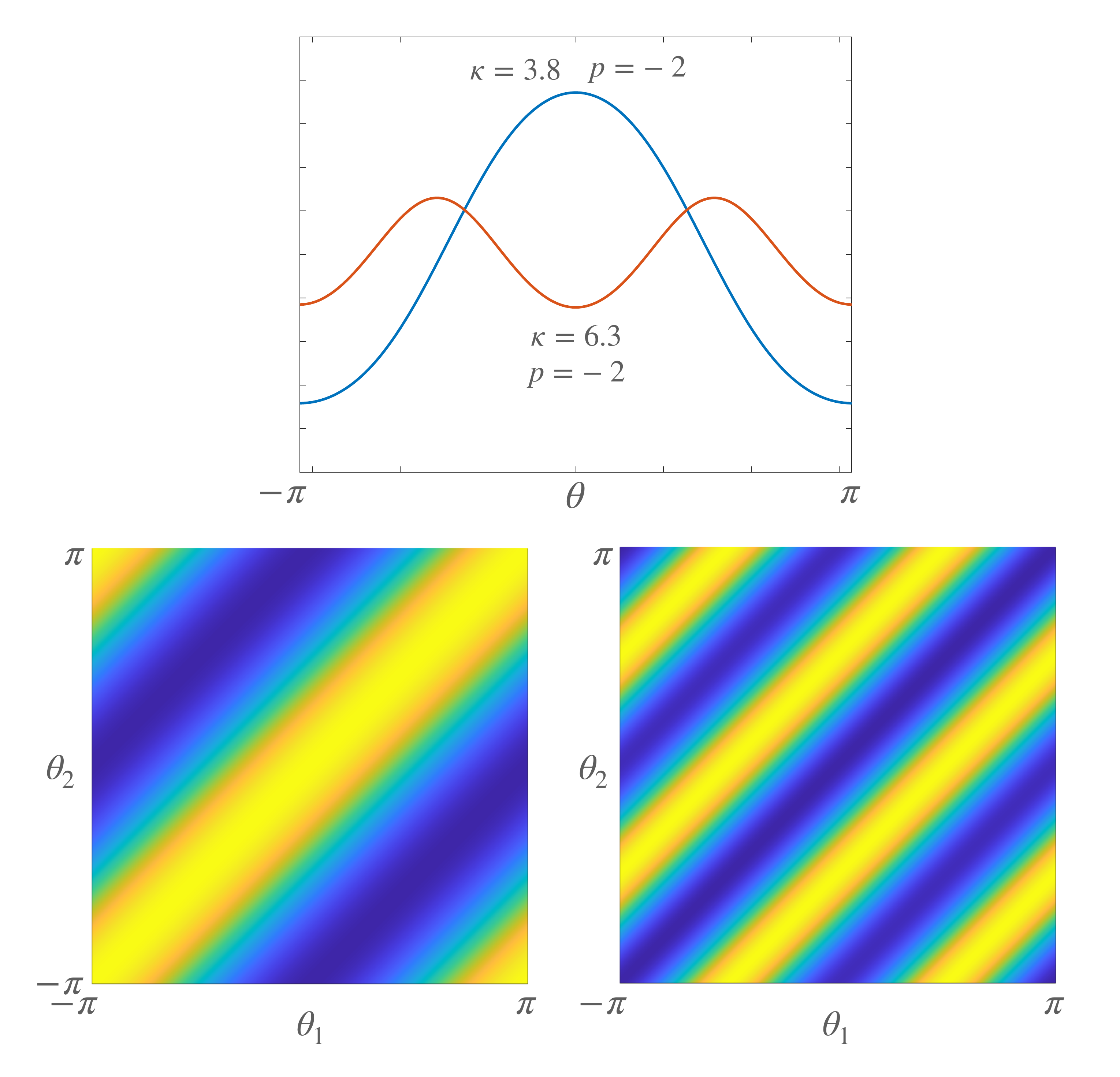}
\caption{An example of the correlated and anti-correlated ground states for two different values of the interaction strength $\kappa=3.8$ and $\kappa=6.3$ with centre of mass momentum $p=-2$. The upper panel shows the ground state probability density as a function of the relative distance. The lower panel shows the corresponding ground state as a function of the two coordinates $\theta_1$ and $\theta_2$ for the two different $\kappa$-values. Yellow indicates high probability and blue low probability.}
\label{correlated}
\end{figure}

In Fig. \ref{energy} we show a plot of the energy as a function of $\kappa$ for different momenta $p$, and for a varying degree of long range interaction, {\it i.e.} for different values of $q$. With a long range gauge potential the ground state is found to have a non-zero centre of mass momentum $p$ provided that the interaction strength $\kappa$ is strong enough. A sufficiently long-range gauge potential is needed in order to have non-zero centre of mass momentum where we in particular note that for the analytically tractable gauge potential described by the short range Dirac delta function the ground state will always have zero momentum.  In addition, the ground state wave function can take two forms, which we refer to as correlated and anti-correlated states, which represent the physical situation where the particles are most likely to be found close to each other (correlated) or separated (anti-correlated), see Fig. \ref{correlated}.  The correlated and anti-correlated states can be expected from looking at the effective potential $V_{eff}(x)$ in Eq. (\ref{poteff}) which is a quartic function of $x$ and consequently can have a single or two minima depending on the parameter $\kappa$ and center of mass momentum $p$. The  ground state, even if it is localised as a function of the relative distance between the particles, carries no information about where in the ring the two particles can be found. In other words, there is an equal probability to measure particle 1 anywhere in the ring, and particle 2 consequently close by or separated depending on whether the state is the correlated or anti-correlated one. This is illustrated in Fig. \ref{correlated} for $q=1$, where the probability density $|\varphi(\theta_1,\theta_2)|^2$ is shown as a function of the two coordinates $\theta_1$ and $\theta_2$.  
 
\section{The role of measurements}

\begin{figure}
\includegraphics[width=7.5cm]{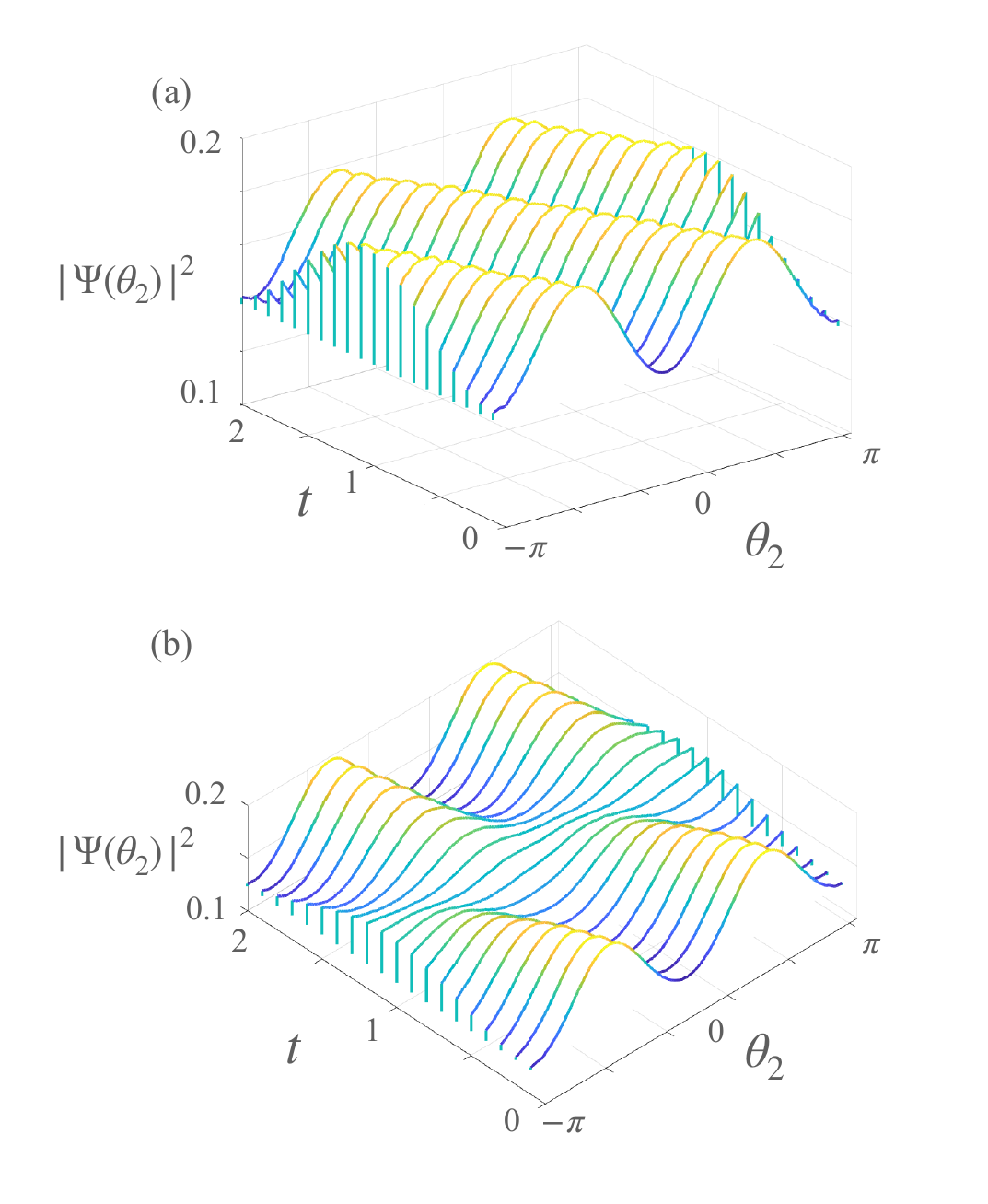}
\caption{Measuring particle 1 with a non-zero uncertainty results in particle 2 propagating around the circle. (a) For a sufficiently large uncertainty in measuring particle 1 with $n=1$, the remaining particle 2 is found in a rotating state with non-zero angular momentum where it retains its shape without any significant spreading. (b) For an almost perfect measurement with a small uncertainty in position of particle 1 with $n=50$, the probability density of particle 2 no longer retains it shape.}
\label{measure}
\end{figure}

From Fig. \ref{correlated} we note that if the position of one particle is measured, then there is an equal probability to find this particle anywhere in the ring. On the other hand, once the first particle is measured, the second particle is most likely to be found close to the first measured particle in the case of a correlated ground state, and separated from it for the anti-correlated state. If we assume a perfect measurement of the position of the first particle, such that it is removed from the system once measured, the resulting wave function for the other particle will be given by the ground state $\Psi(\theta_1^{(0)},\theta_2)$ (see Fig. \ref{correlated}) where $\theta_1^{(0)}$ is the position of the measured particle. After the measurement the remaining particle is no longer in its ground state. Consequently the resulting dynamics is nontrivial where the shape of the wave function and its localised nature can be quickly lost, see Fig. \ref{measure}a. For an imperfect measurement of the first particle, the resulting wave function for the remaining particle is given by 
\begin{equation}
\Psi(\theta_1^{(0)},\theta_2)=\int_{-\pi}^{\pi}d\theta_1 G(\theta_1-\theta_1^{(0)})\Psi(\theta_1,\theta_2),
\end{equation}
where $G(\theta_1-\theta_1^{(0)})$ captures the uncertainty of the measurement and is a distribution which obeys the periodicity of the ring, and with a non-zero width. With increasing uncertainty in position we observe that the wave function for the remaining particle retains its localised shape, and at the same time rotates around the ring due to the non-zero angular momentum of the ground state. This is illustrated in Fig. \ref{measure} where the effect of different uncertainties in position are compared, and where we consider the situation with a non-zero center of mass angular momentum in the ground state. In Fig. \ref{measure} we have chosen a distribution of the form $G((\theta-\theta_1^{(0)})/2))={n!/}{(2\sqrt{\pi}\Gamma[n+\frac{1}{2}]}\cos^{2n}((\theta-\theta_1^{(0)})/2)$ which is normalised to one such that the width is a function of $n$. Similar to the choice of gauge potential in Eq. (\ref{sch1}) the distribution approaches the Dirac delta function in the limit $n$ going to infinity.

For the broadest choice of distribution with $n=1$ in Fig. \ref{measure} the wave function for the now single particle dynamics retains its localised shape for several roundtrips. This may seem counterintuitive at first, since one would expect a localised initial state to expand and eventually interfere with itself in the ring giving rise to complicated dynamics. However, it is possible to construct special solutions to the single particle Schr\"odinger equation for a ring, that are localised non-dispersive solutions with a non-zero angular momentum, {\it i.e.} they rotate. Such solutions have been found by Qin \cite{qin2019} and are of the form
\begin{equation}
\Phi(\theta)=C_0\cos(m\theta-\ell t)e^{i(\ell\theta-\frac{1}{2}(m^2+\ell^2)t)},
\label{scripta}
\end{equation} 
where $m$ and $\ell$ are integers. From Eq. (\ref{scripta}) we notice that $\Phi(\theta)$ is a travelling localised solution with velocity $\ell/m$. In addition, because of the linearity of the single-particle Schr\"odinger equation, we can always construct superpositions of different states such that we have a non-zero background for instance. This is indeed the type of solution we find after measuring one particle with sufficiently large uncertainty, see Fig. \ref{measure}, if we choose $m=\ell=1$ in Eq. (\ref{scripta}). We therefore conclude that when extracting information about the two-particle ground state by measuring the position of one particle with sufficiently large uncertainty, the ground state is projected onto a single particle wave function which is localised, shape preserving, and rotating in the ring. 

\begin{figure}
\includegraphics[width=7.5cm]{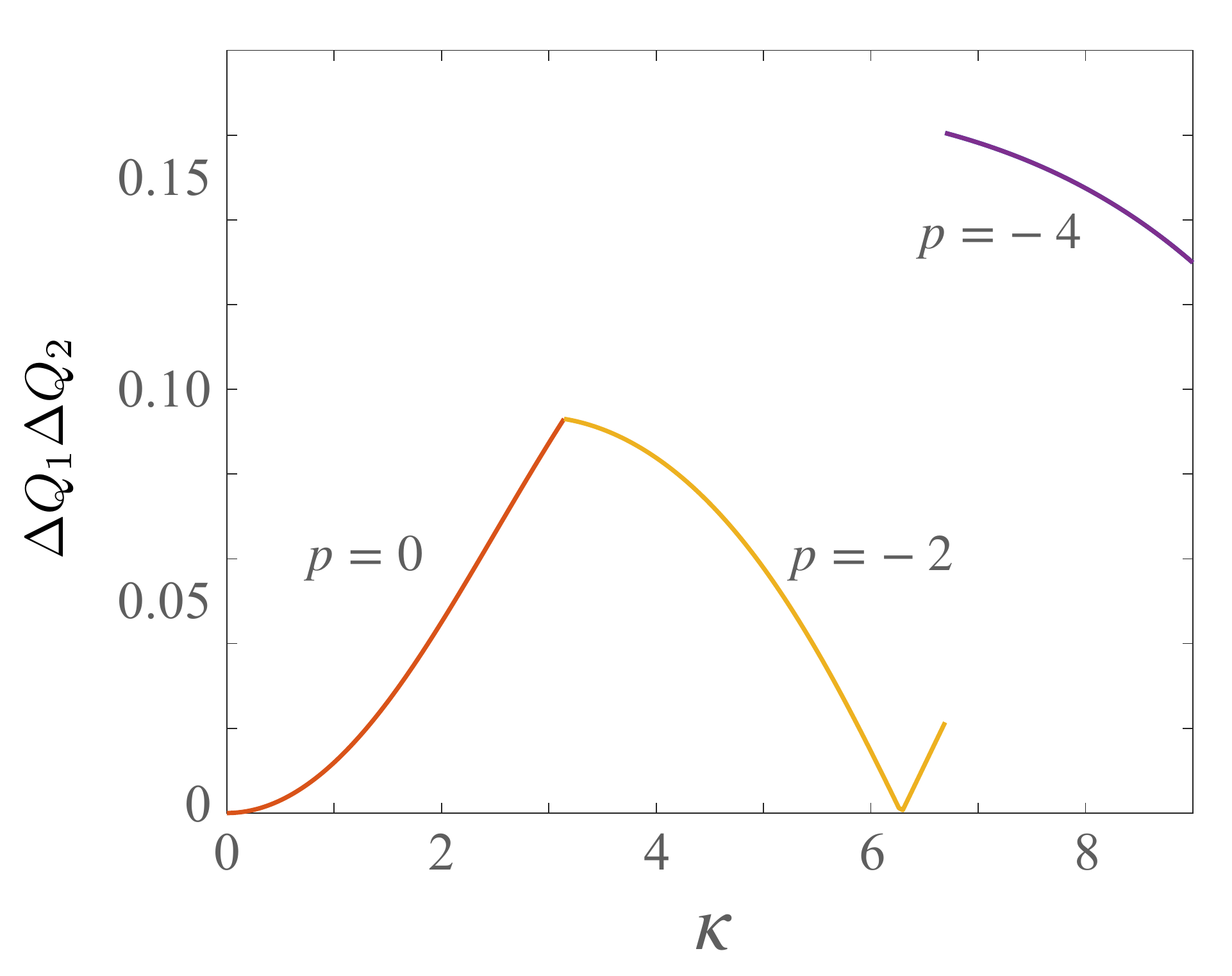}
\caption{The uncertainty $\Delta  Q_1 \Delta Q_2$ for operators defined by Eq. (\ref{oper}) allows us to quantify the correlation in position for the two particles.  We note two special points, $\kappa=\pi$ and $\kappa=2\pi$ which correspond to specific symmetric shapes of the effective potential $V_{eff}(x)$ in Eq. (\ref{poteff}). The corresponding ground state has non-zero centre of mass angular momentum $p$ for $\kappa>\pi$.}
\label{unc}
\end{figure}

The non-dispersive nature of the wave function for the remaining particle after measuring the other particle, leads us naturally to the question of uncertainty relations for the positions of the two particles. We have seen that in order to obtain a non-dispersive wave packet we need an uncertainty in our measurement of the particle 1 position. We can indeed calculate the Robertson-Schr\"odinger uncertainty relation for the positions of particle 1 and particle 2, however, we cannot use the angular coordinate as our position operator with eigenvalues spanning $[-\pi,\pi]$ because of the discontinuous nature which leads to ill-defined expectation values of position and corresponding uncertainties. Instead we define the operator 
\begin{equation}
\hat Q_i=\sum_n e^{i\theta_n}\hat D(R_n)
\label{oper}
\end{equation}
which does not suffer from any discontinuities and is periodic around the ring, and where the operator $\hat D(R_n)$ is a projection operator representing a measurement of a particle in an angular region $R_n$ with eigenvalues $0$ or $1$. With the definition of $\hat Q_i$ we can calculate the corresponding uncertainty relation with respect to $\hat Q_1$ and $\hat Q_2$ with 
\begin{equation}
\Delta  Q_1^2 \Delta Q_2^2 \ge |\frac{1}{2}\langle\{\hat Q_1,\hat Q_2\}\rangle-\langle \hat Q_1\rangle \langle \hat Q_2\rangle |^2+\frac{1}{2i} \langle [\hat Q_1,\hat Q_2]\rangle |^2.
\end{equation}
The uncertainty relation will depend on the quantum state that the two particles are prepared in. In Fig. \ref{unc} we see how the uncertainty  $\Delta  Q_1 \Delta Q_2$  changes drastically as a function of the interaction strength $\kappa$. Two special values $\kappa=\pi$ and  $\kappa=2\pi$ can be identified. At $\kappa=\pi$ the ground state changes between the centre of mass momentum $p=0$ and $p=-2$, with a corresponding change in uncertainty $\Delta  Q_1 \Delta Q_2$ which has a maximum at $\kappa=\pi$. For $\kappa=2\pi$ the uncertainty relation is zero with the centre of mass momentum $p=-2$. We can understand the reason why these two points are special if we look at the effective potential in Eq. (\ref{poteff}). For  $\kappa=\pi$ the $V_{eff}(x)$ provides the same confinement for both $p=0$ and $p=-2$, when we rely on the rotational symmetry of ring with  $V_{eff}(x)=\cos^4(x/2)$ and $V_{eff}(x)=\sin^4(x/2)$ respectively.  

 \section{Conclusions}
The two-body problem in a ring where the interaction is mediated by a gauge potential which in turn depends on the distance between the particles, provides some surprising results. Rotating ground states in combination with a flux penetrating the ring is not uncommon, and has been studied a lot. However, the distance dependent gauge potential gives rise to rotating ground states which are correlated, and, in combination with measuring the position of one of the particles, can result in non-dispersive rotating states for the remaining particle. It is certainly tempting to speculate whether the gauge potential studied in this work could be realised experimentally. Density dependent gauge potentials for a Bose-Einstein condensate in the mean field limit has been studied \cite{edmonds2013a,valenti2020} and also experimentally realised \cite{frolian-2022}, although these works were based on a short range interaction between the atoms. An optical realisation of the two-body problem together with a long-range gauge potential is not inconceivable. Recent advances in the control of light propagation in photonic lattices allows for emulating synthetic nonlinearities which can also be long-range \cite{mukherjee2018,duncan2020}. This, together with a mapping between single particle physics in 2D and interacting 2-particle dynamics in 1D lattices \cite{mukherjee2016} may well provide a route towards an experimental realisation of the phenomena discussed here.

\section{Acknowledgements}
The authors acknowledge helpful discussions with E. Andersson, M. Edmonds and M. Valiente. G.V.-R. acknowledges support from EPSRC CM-CDT Grant No. EP/L015110/1.


\end{document}